\documentclass[12pt]{article}

\usepackage{epsfig}

\usepackage{amssymb}

\renewcommand{\tan}{\mathrm{tg}}

\begin{document}
\title{\bf The beam energy calibration system for the BEPC-II collider}
\author{M.N.~Achasov$^1$, ChengDong Fu$^2$, Xiaohu~Mo$^2$, 
        N.Yu.~Muchnoi$^1$\thanks{e-mail:muchnoi@inp.nsk.su}, \\
        Qing~Qin$^2$, Huamin~Qu$^2$, Yifang~Wang$^2$, Jinqiang~Xu$^2$ \\[2mm]
    \it $^1$G.I.~Budker Institute of Nuclear Physics \\
    \it Siberian Branch of the Russian Academy of Sciences,\\ 
    \it 630090, Novosibirsk, Russia\\[2mm]
    \it $^2$Institute of High Energy Physics, \\
    \it Chinese Academy of Sciences\\
    \it 100049, Beijing, China}
\date{}
\maketitle

%

\begin{abstract}
This document contains a proposal of the BEPC-II collider  beam energy 
calibration system (IHEP, Beijing). The system is based on Compton 
backscattering of carbon dioxide laser radiation, producing a beam of high 
energy photons. Their energy spectrum is then accurately measured by HPGe 
detector. The high-energy spectrum edge will allow to determine the average 
electron or positron beam energy $\varepsilon$ with accuracy 
$\delta\varepsilon/\varepsilon \simeq 3\cdot10^{-5}$.
\end{abstract}

\section{Introduction}

The accurate beam energy calibration of $e^+e^-$ colliders is important for 
particle mass measurements. In the ``$c-\tau$'' energy region the masses of 
$\tau$-lepton, charmonium and charmed mesons are of interest. In some cases 
the energy scale of colliders can be calibrated with extremely high accuracy 
using the resonant depolarization technique \cite{resdep}. But this approach 
is hardly applied for the $e^+e^-$ factories i.e. high luminosity colliders 
with huge electron and positron currents in multi-bunch operation mode. 
The great advance in the luminosity at such colliders are possible due to 
creation of fast bunch-to-bunch feedback systems, that usually have a strong 
depolarization impact on the beam. So at these machines other approaches to 
the beam energy determination should be used.

The project for the BEPC-II \cite{BESIII} beam energy calibration system is 
based on the rather new technique for the beam energy measurement, Compton 
backscattering of monochromatic laser radiation on the electron and positron 
beams. The approach was already experimentally approved at the BESSY-I, 
BESSY-II \cite{BESSY-I}, VEPP-4M and VEPP-3 storage rings \cite{vepp}. 

The general idea is based on the following:
\begin{itemize}
\item The maximal energy of the scattered photon $\omega_{max}$ is strictly 
related with the electron energy $\varepsilon$ due to the kinematics of 
Compton process:
$$\omega_{max}={{\varepsilon^2}\over{\varepsilon+m_e^2/4\omega_0^2}},$$
where $\omega_0$ is the laser photon energy. If one measured $\omega_{max}$, 
then the electron energy could be calculated:
$$\varepsilon={\omega_{max}\over 2} \Biggl[ 1 + 
  \sqrt{1+{ m_e^2 \over {\omega_0\omega_{max}} } }\;\Biggr].$$
\item  
Ultimately high resolution of contemporary commercially available High Purity 
Germanium (HPGe) detectors allow to have the statistical accuracy in the beam 
energy measurement at the level of  
$\delta\varepsilon/\varepsilon \simeq (1\div2)\cdot10^{-5}$.
\item 
The systematical accuracy is mostly defined by absolute calibration of the 
detector energy scale. The accurate calibration could be performed  in the 
photon energy range up to 10~MeV by using of the $\gamma$-active radionuclides.
\end{itemize}

In whole the measurement procedure looks as follows. The laser light is put in
collision with electron or positron beams, the backscattered photons are 
detected by the HPGe detector with ultimate energy resolution. The maximal 
energy of the scattered photons is measured using abrupt edge in the energy 
spectrum. The method was tested using the resonant depolarization technique 
in experiments at VEPP-4M collider \cite{KEDR}. This tests show that two 
methods agree with accuracy  $\Delta\varepsilon=40$~keV, or  
$\Delta\varepsilon/\varepsilon\approx 2\cdot 10^{-5}$.

\section{The Detailed Description of the Approach}
\subsection{The Compton scattering} \label{par:BCS}

Kinematics of photon-electron scattering is defined by the equation:

\begin{equation}
p + k_0 = p' + k\:,
\label{pk1}
\end{equation}
where $p=[\varepsilon,\vec{p}]$, $k_0=[\omega_0,\vec{k_0}]$ are the 
four-momenta of the initial and $p'=[\varepsilon',\vec{p'}]$, 
$k=[\omega,\vec{k}]$ of the final electron and photon respectively. After the 
following transformation of eq. (\ref{pk1}): 

\begin{equation}
(p+k_0-k)^2= p'^2, \mbox{~~~} p k_0 - p k - k_0 k = 0\:,
\label{pk2}
\end{equation}
one can obtain:
\begin{equation}
\varepsilon \omega_0  - |\vec{p}|\omega_0 \cos\alpha -
\varepsilon \omega    + |\vec{p}|\omega   \cos\theta -
\omega \omega_0 (1-\cos\Theta) = 0\:,
\label{pk3}
\end{equation}
 where
\begin{description}
\item[$\alpha$] -- angle between initial electron and initial photon momenta,
\item[$\theta$] -- angle between initial electron and final photon momenta,
\item[$\Theta$] -- angle between initial and final photons momenta.
\end{description}

 Using eq. (\ref{pk3}) the scattered photon energy can be written as follows
\begin{equation}
\omega = \omega_0 \frac{1-\beta\cos\alpha}
{1-\beta\cos\theta+\displaystyle\frac{\omega_0}{\varepsilon}(1-\cos\Theta)}\:, 
\label{w'}
\end{equation}
where $\beta$ is the electron velocity in units of $c$.

When the angle between momenta of initial electron and final photon is equal to
zero ($\theta=0$ and $\Theta=\alpha$) the energy of the scattered photon
becomes  maximal: 
\begin{equation}
\omega_{max} = 
\omega_0 \frac{1-\beta\cos\alpha}{1-\beta+\displaystyle
\frac{\omega_0}{\varepsilon}(1-\cos\alpha)}\:.
\end{equation}
In case of the low-energy photon scattering on the relativistic electron 
($\varepsilon \gg m \gg \omega_0$) the $\omega_{max}$ is
\begin{equation}
\omega_{max} \simeq
\frac{\varepsilon^2 \sin^2\displaystyle\frac{\alpha}{2}+
\displaystyle\frac{m_e^2}{4}\cos\alpha}
{\varepsilon\sin^2\displaystyle\frac{\alpha}{2} + 
\displaystyle\frac{m_e^2}{4\omega_0} }\:.
\label{wr'}
\end{equation}

If $\alpha=\pi$ (head-on collision) then  
\begin{equation}
\omega_{max} 
\simeq 
\frac{\varepsilon^2}{\varepsilon +
\displaystyle\frac{m_e^2}{4\omega_0\sin^2\displaystyle\frac{\alpha}{2}}}=
\varepsilon \frac{\lambda}{1+\lambda}\:,
\label{Xedge}
\end{equation}
where $\lambda=\displaystyle
\frac{4\omega_0\varepsilon}{m_e^2}\sin^2(\frac{\alpha}{2})$.

The total cross section of Compton scattering on unpolarized electron is given
by:
\begin{equation}
\sigma_c=\frac{2\sigma_o}{\lambda}\biggl\{\biggl[1-\frac{4}{\lambda}-
\frac{8}{\lambda^2}\biggr]ln(1+\lambda)+\frac{1}{2}\biggl[1-
\frac{1}{(1+\lambda)^2}\biggr]+\frac{8}{\lambda}\biggr\}\:,
\label{Compton_total_cross_section}
\end{equation}
where $\sigma_o=\pi \cdot r_e^2$ and $r_e$ is a classical electron radius.  If
$\lambda \ll 1$  the cross section is equal to 
\begin{equation}
\sigma_c = \frac{8}{3}\pi r_e^2 (1-\lambda)\:.
\label{Ctcsl}
\end{equation} 

The energy spectrum of scattered photons is:
\begin{equation}
\frac{d\sigma_c}{dy} =
\frac{2\sigma_o}{\lambda}\biggl\{\frac{1}{1-y}+1-y-\frac{4y}{
\lambda(1-y)}+\frac{4y^2}{\lambda^2(1-y)^2}\biggr\}\:,
\label{Compton_energy_spectrum} 
\end{equation}
where $y=\omega/\varepsilon$. 

\begin{figure}[h]
\centering
\fbox{\includegraphics*[width=0.8\textwidth]{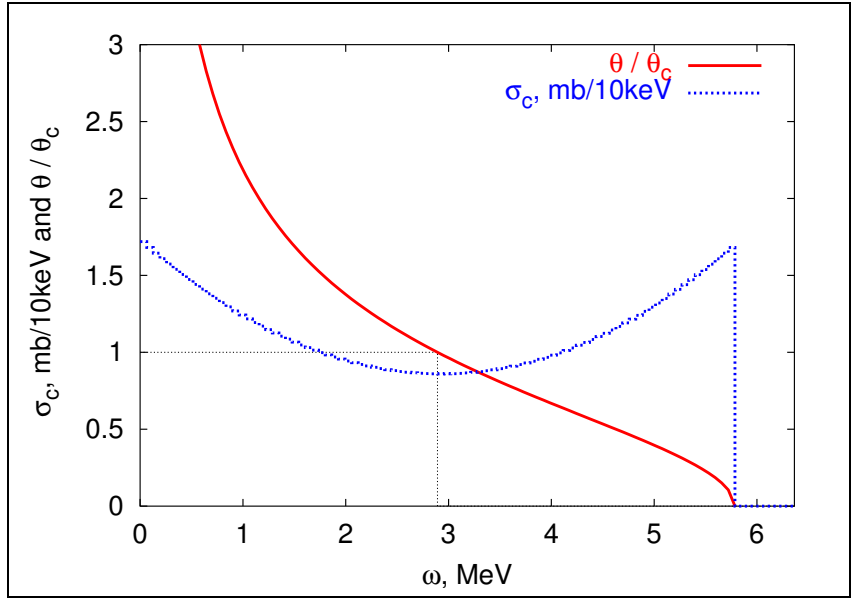}}
\caption{Dotted line is the photon energy spectrum ($\omega_0=0.12$~eV, 
$\varepsilon=1777$~MeV, $\alpha=\pi$). Solid line shows the dependence of the 
$\theta/\theta_c$ on the final photon energy.}
\label{fig:spectrum} 
\end{figure}

Using eq.(\ref{w'}) one can obtain relation between  $\omega$ and $\theta$ 
($\alpha\simeq\pi$): 
\begin{equation}
\omega(\theta) =  
\omega_0 
\frac{1+\beta}{1+\displaystyle\frac{\omega_0}{\varepsilon}+
\cos\theta(\displaystyle\frac{\omega_0}{\varepsilon}-\beta)}
\simeq
\frac{\omega_{max}}{1+\bigl(\displaystyle\frac{\theta}{\theta_c}\bigr)^2}\:,
\label{aec}
\end{equation}
where
\begin{equation}
\theta_c=\displaystyle\frac{m}{\varepsilon}\sqrt{1+\lambda}\:.
\label{thc}
\end{equation}

The differential cross section of the Compton scattering and dependence of 
$\theta/\theta_c$ on $\omega$ is shown in the same plot 
(Fig.~\ref{fig:spectrum}). The edge in the scattered photon energy spectrum 
can be used for the initial electrons energy determination.

\subsection{Measurement of the beam energy}\label{e-measurement}

The Compton scattering can be used for collider beam energy calibration. The 
initial electron energy $\varepsilon$ is related with $\omega_{max}$ as
\begin{equation}
\varepsilon = 
\frac{\omega_{max}}{2}\Biggl(1 + 
\sqrt{1+
\frac{m^2}{\omega_0\omega_{max}\sin^2\displaystyle\frac{\alpha}{2}}}\Biggr)
\simeq
\frac{m}{2\sin\displaystyle\frac{\alpha}{2}}\sqrt{\frac{\omega_{max}}{\omega_0}}\:.
\label{E}
\end{equation}

\begin{figure}[htbp]
\centering
\includegraphics*[width=0.8\textwidth]{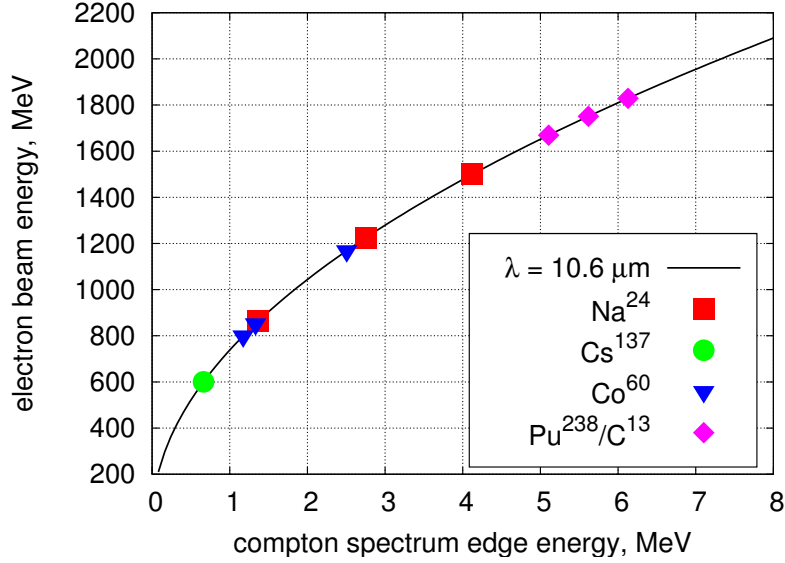}
\caption{Relation between $\omega_{max}$ and $\varepsilon$ (solid line).  Dots
is the energies of $\gamma$-active radionuclides (reference lines for HPGe 
detector calibration).}
\label{fig:isotopes}
\end{figure}

It is possible to obtain the initial electron energy by measuring the maximal
energy of the scattered  photon (Fig.~\ref{fig:isotopes}).
The monochromatic laser radiation is a proper source of initial photons with
$\omega_0\simeq0.12$~eV. In this case the scattered photons energy 
$\omega_{max}$ is about 1 -- 10 MeV. 

The energy of the photons can be precisely measured by HPGe. The detector 
energy scale can be accurately calibrated in this energy range  by using the 
well-known radiative sources of $\gamma$-radiation (Fig.\ref{fig:isotopes}).

Further we will understand $\varepsilon$ and $\omega_0$ as the average energy 
of the electron and laser beam respectively. 

\begin{figure}[h]
\centering
\includegraphics*[width=0.85\textwidth]{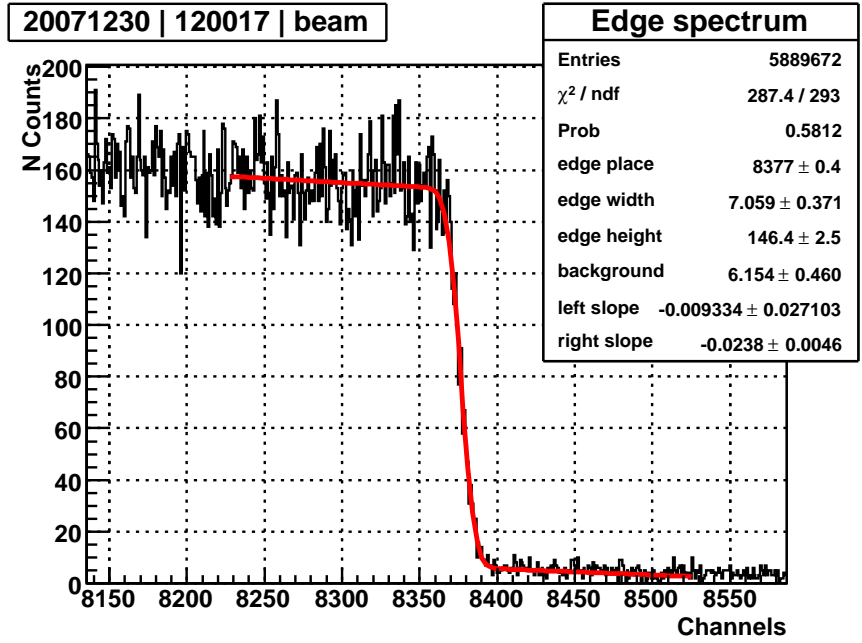}
\caption{The measured edge of the scattered photons energy spectrum. The line 
is the fit result.}
\label{fitspec}
\end{figure}

The $\omega_{max}$ is measured by fitting of the energy spectrum edge by the 
erfc-like function (Fig.\ref{fitspec}):
\begin{eqnarray}
 g(x,p_{0{\ldots} 5}) = {1\over2}\bigl(p_4(x-p_0)+p_2)\bigr) \times 
 erfc\Biggl[ { {x-p_0} \over {\sqrt{2}p_1} }\Biggr] - \nonumber \\
 -{ {p_1 p_4} \over {\sqrt{2\pi}} } \times 
 exp\Biggl[- {{(x-p_0)^2}\over{2p_1^2}}\Biggr]  + p_5(x-p_0)+p_3,
\end{eqnarray}
where $p_0$ is the edge position, $p_1$ is the edge width, $p_2$ is the edge
amplitude, $p_3$  is background parameter, $p_4$ and $p_5$ describe slopes at
right and left from the edge.

Under the actual experimental conditions the abrupt edge of the scattered 
photons energy distribution (Fig.\ref{fitspec}) is smeared due to the 
following effects:
\begin{itemize}
\item the energy spread in the electron beam $\delta\varepsilon/\varepsilon$, 
\item the width of laser radiation line $\delta\omega_0/\omega_0$,
\item the energy resolution of the HPGe detector $\delta r/r$, 
\item the angular distribution between initial particles. 
\end{itemize}
Taking into account these contributions the visible edge width can be written
as
\begin{equation}
\sigma_{\omega} \equiv
\frac{\delta\omega_{max}}{\omega_{max}} 
\simeq
2\frac{\delta\varepsilon}{\varepsilon}
\oplus
\frac{\delta\omega_0}{\omega_0}
\oplus
\frac{\delta r}{ r}
\oplus
\frac{\delta\alpha}{\tan\displaystyle\frac{\alpha}{2}}\:,
\label{dwda}
\end{equation}
where $\delta\alpha \simeq \sigma'_e \oplus \sigma'_l$  is the square sum of
the electron and laser beams angular spreads. This contribution is negligible 
when $\alpha\simeq\pi$.

The relative accuracy of $\varepsilon$ measurement is
\begin{equation}
\frac{\Delta\varepsilon}{\varepsilon} \simeq 
\frac{1}{2} \frac{\Delta\alpha}{\tan(\alpha /2)}
\oplus
\frac{1}{2} \frac{\Delta\omega_0}{\omega_0}
\oplus
\frac{1}{2} \frac{\Delta\omega_{max}}{\omega_{max}}\;.
\label{deda}
\end{equation}
Here the contribution due to the angle $\alpha$ uncertainly can be omitted 
also. The contribution of the laser photon energy stability is negligible 
$\Delta\omega_0/\omega_0 \lesssim 10^{-7}$ (see 
section~\ref{sec:laser} below). The last term is accuracy of the backscattered
photons energy measurement. It depends on the width of the spectrum edge, 
events number and the accuracy of the HPGe detector absolute calibration.  
The statistical accuracy of the $\omega_{max}$ determination can be estimated 
as follows:
\begin{equation}
\Delta\omega_{max}
\simeq \sqrt{
\frac{2 \delta\omega_{max}}
{\displaystyle\frac{dN_\gamma}{d\omega}\bigl(\omega_{max}\bigr)}
}\;,
\label{Dedge}
\end{equation}
where $\delta\omega_{max}$ is the edge width eq.~(\ref{dwda}) and 
$\displaystyle\frac{dN_\gamma}{d\omega}\bigl(\omega_{max}\bigr)$ is the photon
density at the spectrum edge. The density can be estimated as:
\begin{equation}
\frac{dN_\gamma}{d\omega}\bigl(\omega_{max}\bigr) =
\epsilon \cdot L_{e\gamma} \cdot t \cdot 
\frac{d\sigma_c}{d\omega}\bigl(\omega_{max}\bigr)\;,
\label{nedge}
\end{equation}
where $\displaystyle\frac{d\sigma_c}{d\omega}\bigl(\omega_{max}\bigr)$ is the 
Compton cross section at the edge  eq.~(\ref{Compton_energy_spectrum}), 
$\epsilon$ is the photons detection efficiency, $L_{e\gamma}$ is the 
photon-electron luminosity and $t$ is the spectrum accumulation time. 

\subsection{Laser-Electron Luminosity}\label{luminosity}

The electron-photon luminosity $L_{e\gamma}$ define a number of scattered 
photons per one electron bunch:
\begin{equation}
N_\gamma=L_{e\gamma}\cdot\sigma_c(\omega_0,\varepsilon)\;,
\label{lum}
\end{equation}
where  $\sigma_c(\omega_0,\varepsilon)$  is the total cross section of Compton
scattering. The luminosity $L_{e\gamma}$ depends on the intensity of the 
electron and photon beams and their effective cross section.

The laser light propagation can be considered in the framework of the Gaussian 
beam optics approach. By definition, the transverse profile of the intensity 
of the Gaussian beam with a power $P$ can be described with a function 
\cite{rp-photonics}:

\begin{equation}
I(r,s)=\displaystyle\frac{P}{\pi w(s)^2 / 2}
\exp\Biggl\{-2\displaystyle\frac{r^2}{w(s)^2}\Biggr\}\;,
\label{gbp}
\end{equation}
where the beam radius $w(s)$ is the distance from the beam axis where the 
intensity drops to the level of $1/e^2$. The beam radius in free space varies 
along the propagation direction according to:
\begin{equation}
w(s)=w_0 \cdot \sqrt{1+
\Biggl(\displaystyle\frac{\lambda s}{\pi w_0^2}\Biggr)^2}\;,
\label{gbs}
\end{equation}
where $w_0=w(s=0)$ -- beam radius at the beam waist. The radius of curvature 
$R$ of the wave fronts evolves according to:
\begin{equation}
R(s)=s\cdot \Biggl[ 1+\Biggl(\displaystyle\frac{\pi w_0^2}{\lambda s}\Biggr)^2 
\Biggr]\;.
\label{gbr}
\end{equation}

The Gaussian beam state at a certain position $s$ can be specified with a 
complex $q$ parameter:
\begin{equation}
\displaystyle\frac{1}{q(s)} = \displaystyle\frac{1}{R(s)} + 
\displaystyle\frac{i\lambda}{\pi w(s)^2}\;.
\label{gbq}
\end{equation}
The Gaussian beam pass through an optical elements can be described by 
transforming $q$ parameter with an $ABCD$ matrix of the optical system 
\begin{equation}
q' = \displaystyle\frac{Aq+B}{Cq+D}\;.
\label{gbabcd}
\end{equation}
The laser beam transverse photon density is written as
\begin{equation}
\rho_{ph}(x,y,s)= \displaystyle\frac{n_{ph}}{2\pi\sigma(s)^2}
\exp \biggr\{ -\displaystyle\frac{x^2}{2\sigma(s)^2} - 
\frac{y^2}{2\sigma(s)^2} \biggl \}\;,
\label{gbd}
\end{equation}
where \fbox{$\sigma(s) \equiv \displaystyle\frac{1}{2} w(s)$} and $n_{ph}$ is 
the longitudinal density of laser photons. For any laser, operating in CW mode:
\begin{equation}
n_{ph}=\frac{dN_{ph}}{ds}=
\frac{P}{\omega_0 c e} \equiv \frac{P \lambda}{h c^2}\;,
\label{nph}
\end{equation}
where $P$ is the laser CW power,  $\omega_0$ is the laser photon energy, 
$\lambda$ is the laser radiation wavelength. 

The specification table of laser parameters usually contains output beam size 
and divergence given in the form:
\begin{itemize}
\item $S_o$ is the beam size at $1/e^2$ diameter at exit aperture; 
\item $D_o$ is the beam divergence at $1/e^2$ diameter in the far field.
\end{itemize}
First order derivation of eq.~(\ref{gbs}) over $s$ at $s \rightarrow \infty$ 
gives
the Gaussian beam divergence in far field:
\begin{equation}
\frac{1}{2} D_o = \frac{\lambda}{\pi w_0}\:,
\label{bdaff}
\end{equation} 
and the waist size inside laser cavity ($w_0$ -- radius at $1/e^2$ intensity
level) is given by:
\begin{equation}
w_0=\frac{2 \lambda}{\pi D_o}\:.
\label{wsic}
\end{equation} 
Using eq.~(\ref{gbs}) the distance $s_0$ between exit aperture and the waist 
position inside laser cavity can be calculated as follows:
\begin{equation}
s_0 = \frac{1}{D_o^2}\sqrt{S_o^2 D_o^2-(4\lambda/\pi)^2}\:, \mbox{~}
\Biggl(\frac{1}{2}S_o = w_0 \sqrt{1+\biggl(\frac{\lambda s_0}{\pi w_0^2}
\biggr)^2}\Biggr)\:.
\label{solequ1}
\end{equation} 
Using the values of $s_0$, $w_0$, $\lambda$ the initial complex $q$ parameter
can be obtained and traced through the optic system.

The electron beam parameters are:
\begin{itemize}
\item $N_e$ -- number of particles in a bunch
\item $\sigma_x(s) = \sqrt{\epsilon_x \beta_x(s)}$ -- horizontal betatron size 
\item $\sigma_y(s) = \sqrt{\epsilon_y \beta_y(s)}$ -- vertical betatron size
\item $\sigma_e$ -- longitudinal electron beam size
\item $\sigma_\varepsilon$ -- beam energy spread
\item $\varepsilon, \varepsilon_0$ -- particle energy and average particle 
energy
\item $(\varepsilon - \varepsilon_0)/\varepsilon_0\cdot\psi(s)$ -- shift in
$x$ due to dispersion function $\psi$
\end{itemize}
The electron beam distribution can be written as follows:
\begin{eqnarray}
\nonumber & \rho_e(x,y,\varepsilon,s) = 
\displaystyle\frac{N_e}{(2\pi)^{3/2}\sigma_x\sigma_y\sigma_{\varepsilon}} 
\times & \\
& \exp \biggr\{ 
-\displaystyle\frac{(\varepsilon - \varepsilon_0)^2}{2\sigma_\varepsilon^2} 
-\displaystyle\frac{(y           - y_0)^2          }{2\sigma_y^2    } 
-\displaystyle\frac{\bigl(x-x_0 - \psi\displaystyle\frac{\varepsilon - 
\varepsilon_0}{\varepsilon_0}\bigr)^2 }{2\sigma_x^2} 
\biggl\}.
\label{edense}
\end{eqnarray}
Here $x_0$ and $y_0$ depends on $s$ and indicate the possible bias between the
axes of the electron and laser beams. This shift can be a result of the beams 
improper alignment or non-zero crossing angle. 

The beams interact in the limited area with length $\mathcal{L}$. Let us note 
that the electron beam is much shorter than the laser one. The luminosity is 
\begin{eqnarray}
{{dL_{e\gamma}}\over{dsd\varepsilon}} = 
\int\int 2 \cdot \rho_{ph}(x,y,s)\rho_e(x,y,s) dxdy
\end{eqnarray}
Here factor 2  is called M{\o}ller luminosity factor \cite{moller}. After 
integration the luminosity is
\begin{eqnarray}
\nonumber  & \displaystyle\frac{dL_{e\gamma}}{dsd\varepsilon}=
\frac{2 n_{ph} N_e}
{(2\pi)^{3/2} \sigma_\varepsilon \sqrt{\bigl(\sigma^2+\sigma_x^2\bigr) 
\bigl(\sigma^2+\sigma_y^2\bigr)}}\times & \\
\nonumber & \times \exp \biggr\{-\displaystyle\frac{y_0^2}{2(\sigma^2+
\sigma_y^2)} \biggl\}\times & \\
&\times \exp\biggr\{
- \displaystyle\frac{(\varepsilon - \varepsilon_0)^2}{2\sigma_\varepsilon^2}
- \displaystyle\frac{(x_0 + \psi\displaystyle\frac{\varepsilon 
- \varepsilon_0}{\varepsilon_0})^2}{2(\sigma^2+\sigma_x^2)} 
\biggl\} .
\label{dLdsdE}
\end{eqnarray}
If $x_0 \neq 0$ and $\psi \neq 0$ than the average energy 
$\overline{\varepsilon_L}$ of the electrons which interacts with the laser 
photons  differs from  the average electron beam energy:
\begin{equation}
\Delta\varepsilon = (\overline{\varepsilon_L}-\varepsilon_0) = 
-\displaystyle\frac{x_0\,\sigma_\psi\,\sigma_\varepsilon}{\sigma^2
+\sigma_x^2+\sigma_\psi^2}\;,
\label{deltae}
\end{equation}
where $\sigma_\psi=\psi\cdot(\sigma_\varepsilon/\varepsilon_0)$ is the part of
the transverse horizontal electron beam size, which origins from non-zero 
dispersion function $\psi$. The energy dispersion of interactive electrons 
deviates from the beam energy spread when $\psi \neq 0$:
\begin{equation}
\sigma'_\varepsilon = 
\sigma_\varepsilon \sqrt{\displaystyle\frac{1}{1
+\displaystyle\frac{\sigma_\psi^2}{\sigma^2+\sigma_x^2}}}.
\label{spreadtr}
\end{equation}

The luminosity integrated over $\varepsilon$ is
\begin{eqnarray}
\nonumber & \displaystyle\frac{dL_{e\gamma}}{ds}=
\displaystyle\frac{n_{ph} N_e}
{\pi \sqrt{\bigl(\sigma^2+\sigma_x^2+\sigma_\psi^2\bigr) 
\bigl(\sigma^2+\sigma_y^2\bigr)}}\times & \\
& \times \exp \biggr\{-\displaystyle\frac{y_0^2}{2(\sigma^2+\sigma_y^2)} 
\biggl\}\times 
\exp \biggr\{-\displaystyle\frac{x_0^2}{2(\sigma^2+\sigma_x^2+\sigma_\psi^2)} 
\biggl\}&.
\label{LCW}
\end{eqnarray}

The integration of eq.~(\ref{LCW}) over $s$ (formal limits of integration are
defined by $\mathcal{L}$) could be performed numerically, using actual
electron and laser beams size.

\subsection{CO$_2$ laser as a source of reference energy 
radiation}\label{sec:laser}

The most convenient source of photons for the beam energy calibration system
of the BEPC-II is CO$_2$ laser with radiation wavelength
$\lambda\simeq 10.6\mu$m
(Fig.~\ref{fig:isotopes}). The main requirements to the device are 
\begin{itemize}
\item
the single generated line,
\item 
high stability of parameters,
\item
easy maintenance. 
\end{itemize}
It should be rather exclusive equipment especially to satisfy the first item.

The Carbon dioxide laser is a laser based on a gas mixture as the gain medium,
which contains carbon dioxide (CO$_2$), helium (He), nitrogen (N$_2$), and 
possibly some hydrogen (H$_2$), water vapor, and/or xenon (Xe). Such a laser 
is electrically pumped via a gas discharge, which can be operated with DC 
current, with AC current (e.g. 20-50 kHz) or in the radio frequency (RF) 
domain. Nitrogen molecules are excited by the discharge into a metastable 
vibrational level and transfer their excitation energy to the CO$_2$ molecules
when colliding with them. Helium serves to depopulate the lower laser level 
and to remove the heat. Other constituents such as hydrogen or water vapor can
help (particularly in sealed tube lasers) to reoxidize carbon monoxide (formed
in the discharge) to carbon dioxide.

\begin{figure}[htbp]
\centering
\includegraphics*[width=0.8\textwidth]{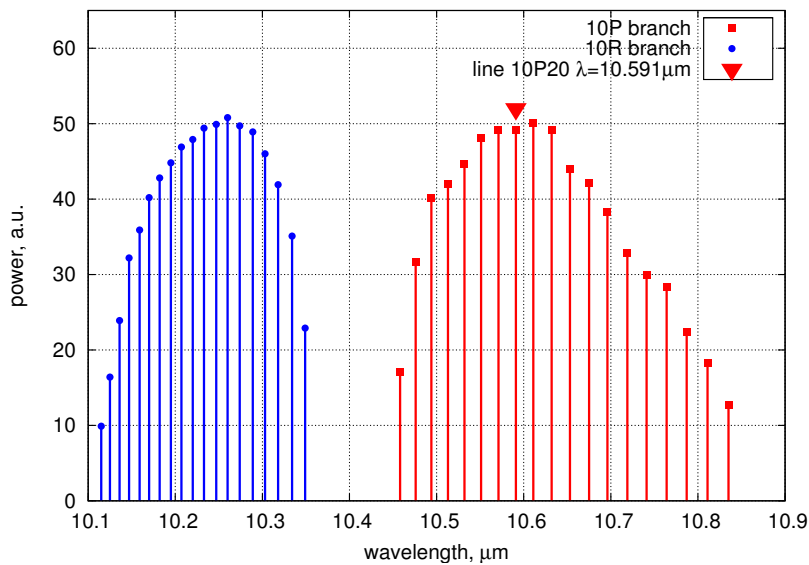}
\caption{Different lines in carbon dioxide laser radiation spectrum (10P and 
10R branches)}
\label{fig:co2lines}
\end{figure}

CO$_2$ lasers typically emit photons with $\lambda\simeq10.6\mu$m, but there 
are other lines in the region of 9-11 $\mu$m (see Fig.~\ref{fig:co2lines}). In
common case laser can radiate photons spontaneously at different neighbor 
lines, or even at several lines simultaneously. As the relative distance 
between neighbor lines $\Delta\nu/\nu \simeq 0.2$~\% (improper for 
calibration), it is needed to have laser operating at only one 
vibration-rotation transition in CO$_2$ molecule. The width of each line is 
formed by Doppler and collision widening, resulting in about 100~MHz bandwidth
($\Delta\nu/\nu \simeq 3$~ppm). 

Averaging the radiation wavelength over longitudinal cavity modes results in 
average laser photon energy stability at level of 
$\Delta\nu/\nu \lesssim 0.1$~ppm. Thus the single-line carbon dioxide laser is
a source radiation adequate to the calibration system.

The GEM Selected 50$^{\mbox{\scriptsize TM}}$ laser produced by the COHERENT 
{\em Inc.} (USA) is the most suitable for designed system. It provides 
25 -- 50 W of CW power at the wavelength 10.591 $\mu$m. The photons average 
energy stability is about 0.1 ppm. The laser of this type is used in the 
similar system at the VEPP-4M collider in Novosibirsk since 2005. During this 
period  no degradation of its parameters were noticed. The laser consumes 
about 1~kW power and needs appropriate cooling system.

\subsection{HPGe semiconductor detector}

HPGe detector is a large germanium diode of the p-n or p-i-n type operated in 
the reverse bias mode. At a suitable operating temperature (normally 
$\simeq$85 K), the barrier created at the junction reduces the leakage current
to acceptably low values. Thus an electric field can be applied that is 
sufficient to collect the charge carriers liberated by the ionizing radiation. 

The energy lost by ionizing radiation in semiconductor detectors ultimately 
results in the creation of electron-hole pairs. The average energy  $\xi$ 
necessary to create an electron-hole pair in a given semiconductor at a given 
temperature is independent of the type and the energy of the ionizing 
radiation. The value $\xi$ is 2.95 eV in germanium at 80 K. Since the 
forbidden bandgap value is 0.73 eV for germanium at 80 K, it is clear that not
all the energy of the ionizing radiation is spent in breaking covalent bonds. 
Some of it is ultimately released to the lattice in the form of phonons.

The constant value of $\xi$ for different types of radiation and for different
energies contributes to the versatility and flexibility of
semiconductor detectors for use in nuclear spectroscopy. The low value of 
$\xi$ compared with the average energy necessary to
create an electron-ion pair in a gas (typically 15 to 30 eV) results in the 
superior spectroscopic performance of semiconductor detectors.

The most suitable HPGe detectors for the 1--10~MeV photons are Coaxial type 
detectors. They have different volumes, and hence, different registration 
efficiency. The efficiency of each detector is usually specified by a 
parameter called {\em relative detection efficiency}. The {\em relative 
detection efficiency} of coaxial germanium detectors is defined at 1.33 MeV 
relative to that of a standard 3-in.-diameter, 3-in.-long NaI(Tl) 
scintillator. 

For the BEPC-II energy calibration system the coaxial HPGe detector 
(50~mm diameter, 50~mm height) with 25\% relative efficiency and 5\% total 
absorption efficiency for 6~MeV photons is relevant. The typical energy 
resolution of such a detector is $\sigma_\omega \simeq$~1~keV for 
$\omega$=1.33~MeV and $\sigma_\omega \simeq$~2.5~keV for $\omega$=6~MeV.

The detector is connected to the spectrometric station, which translates data 
to the USB port of the computer. As far as detector operates at cryogenic 
temperature, it requires the cooling system.

\section{BEPC-II Beam Energy Calibration System}
\begin{figure}[p]
\centering
\includegraphics*[width=\textwidth]{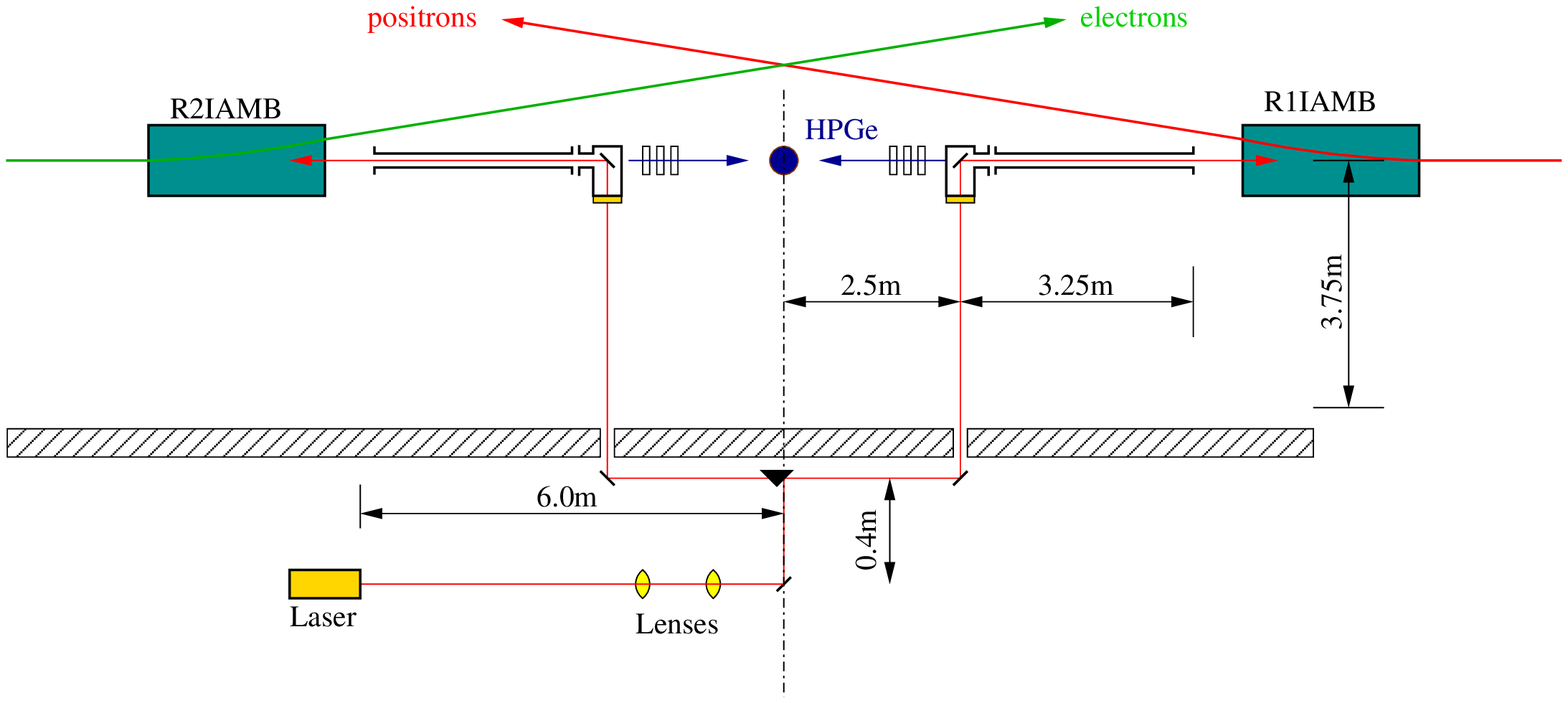}
\caption{The apparatus layout at the BEPC-II.}
\label{fig:layout}
\vspace{1cm}
\includegraphics*[width=\textwidth]{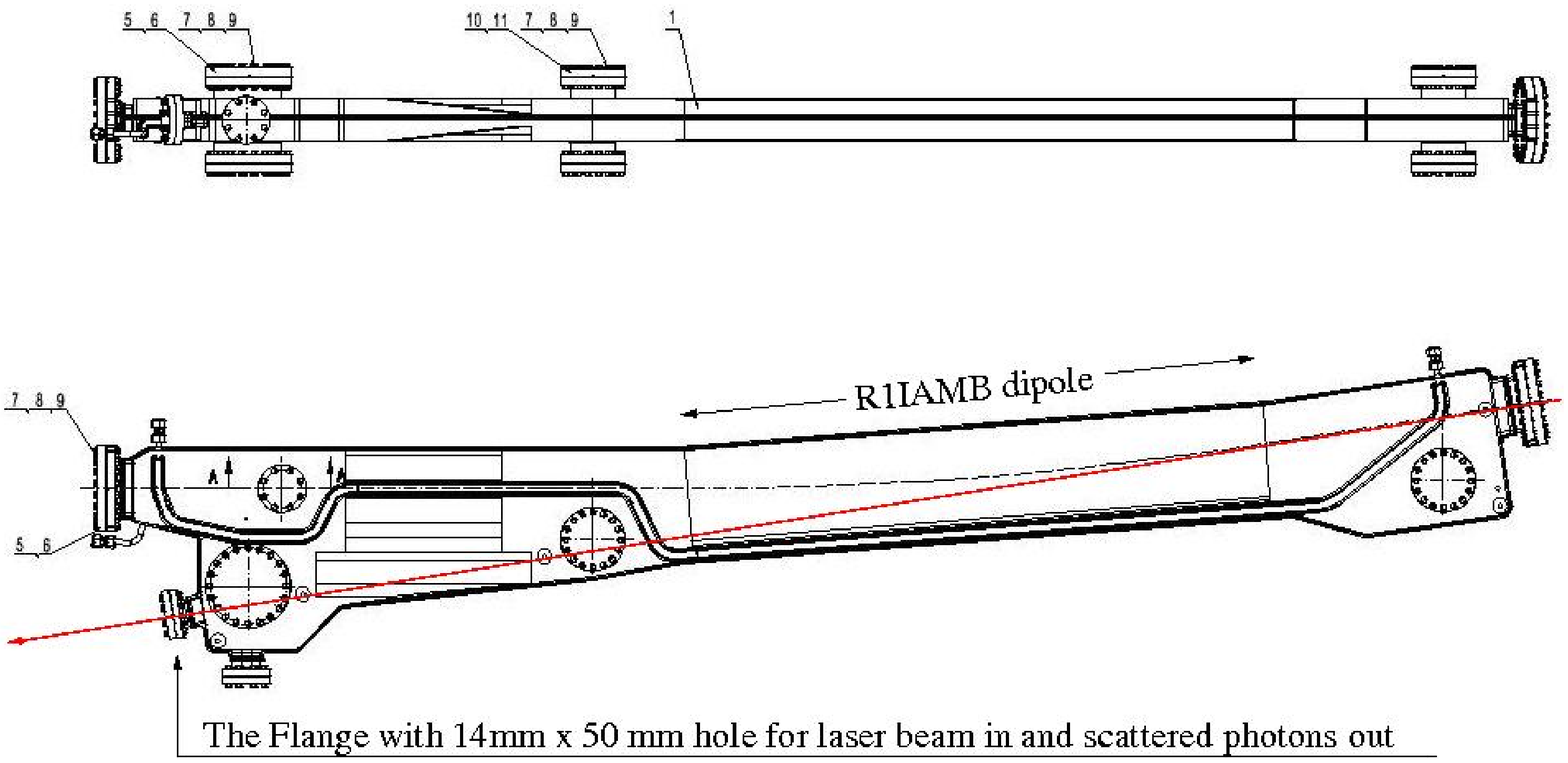}
\caption{BEPC-II vacuum chamber}
\label{fig:vc-1}
\end{figure}

In Ref.\cite{ihep-proposal} it was suggested to calibrate both $e^-$ and $e^+$
beam energies using the same HPGe detector. This approach reduce the cost of 
the system. The system should be placed at the opposite to the beams 
interaction region side of the collider. The BEPC-II beam energy calibration 
system layout is shown in Fig.~\ref{fig:layout} (all the lengths are 
approximate).

The input of the laser radiation into the BEPC-II vacuum chamber is the most 
complex unit of the system. The BEPC-II vacuum chamber design is shown in  
Fig.\ref{fig:vc-1}. The laser beam insertion system is connected to the flange
of the chamber. The distance between the vacuum chambers flanges of the 
electron and positron rings is equal to 11.5 m. The HPGe detector is placed 
between the flanges on the equal distance from both (Fig.~\ref{fig:layout}). 

\subsection{Laser-to-vacuum insertion assembly}

\begin{figure}[h]
\centering
\includegraphics*[width=0.6\textwidth]{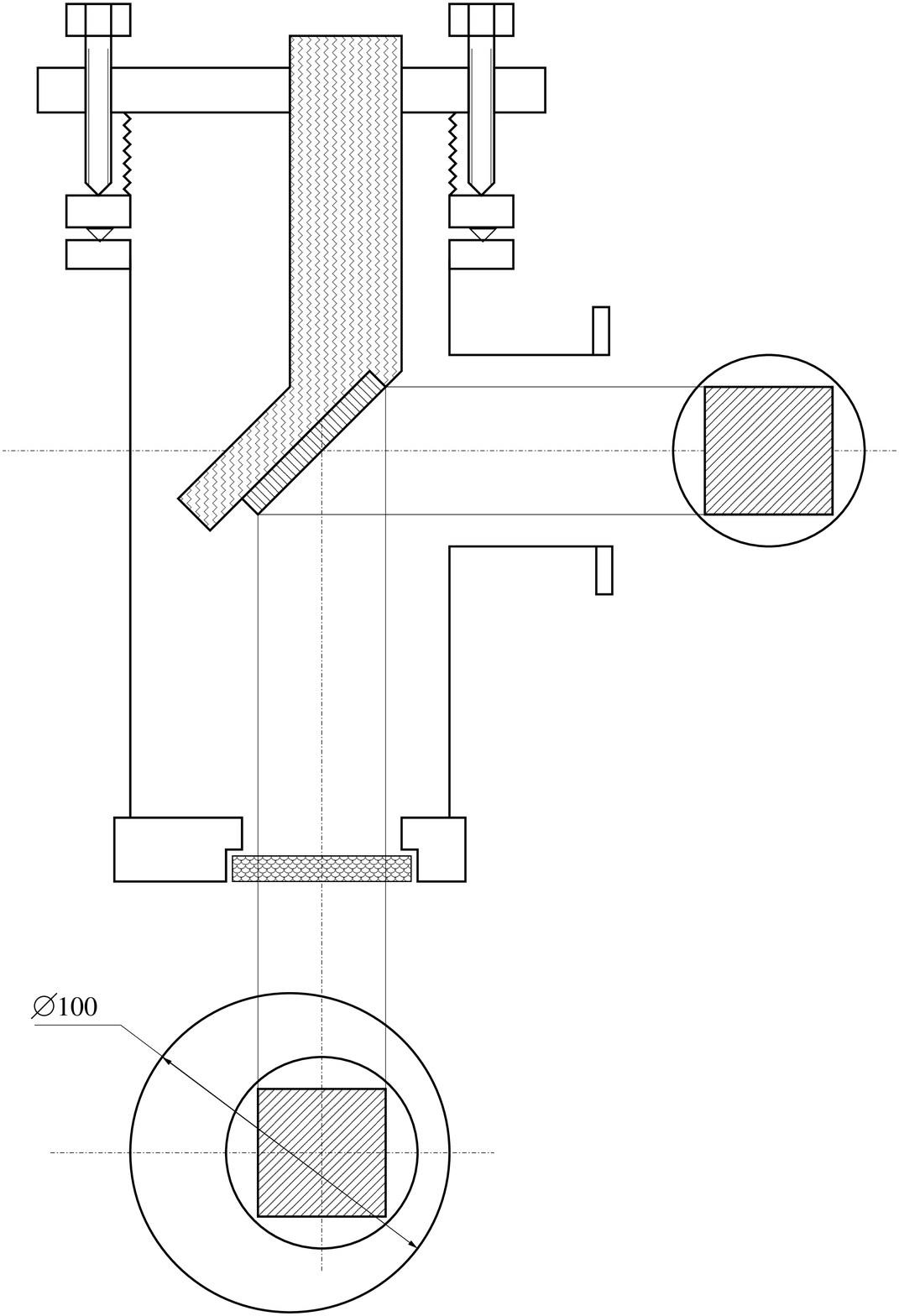}
\caption{Laser-to-vacuum insertion assembly}
\label{fig:vvod}
\end{figure}

The conceptual scheme of laser-to-vacuum insertion system is shown in 
Fig.\ref{fig:vvod}. 

The laser beam is inserted to vacuum chamber through the ZnSe or ZnS 
(Cleartan\textregistered) entrance window. In the  vacuum chamber the laser 
beam is reflected on the angle 90$^\circ$ by a copper mirror, which is mounted
with a good thermal contact on the massive copper support. This support can be
turned by bending the vacuum flexible bellow (sylphon). So the angle between 
the mirror and the laser beam can be adjusted as necessary. After 
backscattering the photons return to the mirror, pass through it and leave the
vacuum chamber. Then they can be detected by the HPGe detector. The SR 
radiation photons fall on the mirror and heat it. In order to reduce the heat 
of the mirror it is placed at the 3.25 m distance from the BEPC-II vacuum 
chamber flange.

\begin{table}[h]
\centering
\begin{tabular}{l|l}
Beam energy  $E$, GeV  & 2 \\
Beam current $I$, A & 0.91 \\
Bending radius  $\rho$, cm & 913 \\
Mirror size $D$, cm & 4.0 \\
Distance to mirror $L=L_1+L_2$, cm & 605 = 280 + 325 \\
$\phi \simeq D/L$, mrad & 6.6 \\
\hline
SR power on the mirror, W & 150 \\
\end{tabular}
\caption{The parameters necessary for calculation of the SR power. $L_1$ is 
the distance from SR photons radiation point to the BEPC-II vacuum chamber 
flange, $L_2$ is the distance from the BEPC-II vacuum chamber 
flange to the mirror.}
\label{tabheat}
\end{table}

The SR radiation power absorbed by the mirror can be estimated as follows:
\begin{equation}
P\mbox{~[W]} = U_0\mbox{~[eV]} \cdot I\mbox{~[A]} \cdot \frac{\phi}{2\pi}\;,
\label{power}
\end{equation}
where $I$ is the beam current, $U_0$ is the particle energy loss per turn due 
to SR radiation, $\phi\simeq D/L$. Here $D\simeq l/\sqrt{2}$ is the mirror 
size seen from the SR photons radiation point ($l$ is the mirror width), $L$ 
is the distance between the mirror and the SR photons radiation point. The 
energy loss $U_0$ can be obtained using the following formula: 
\begin{equation}
U_0=\frac{4\pi}{3} \cdot \alpha \cdot \hbar c \cdot \frac{\gamma^4}{\rho}\;,
\label{Uo}
\end{equation}
where $\rho$ is the beam trajectory radius in the bending magnet. Using the 
values of BEPC-II parameters presented in Table~\ref{tabheat} we obtained 
$P=150$ W. The abstraction of heat outside the vacuum chamber to some cooling 
system is provided by the massive cooper support of the mirror.  

\subsection{Laser -- electron beam interaction area.}

The laser and $e^-$/$e^+$ beams will interact in the straight sections of the 
BEPC-II, enclosed between R1IMB01 and R1IAMB  dipoles for $e^+$ (R2IMB01 and 
R2IAMB for $e^-$). The lattice functions of the BEPC-II straight section are 
shown in Fig.~\ref{fig:lf}. The electron and laser beam size in $x$ and $y$ 
projections versus $z$-coordinate of the beam axis are shown in 
Fig.~\ref{fig:x},~\ref{fig:y}. The laser beam is focused at the BEPC-II vacuum
chamber entrance flange, where the geometrical aperture is minimal (vertical 
size is 14~mm). The divergence of the laser beam is defined by its transverse 
size at the waist, see eq.(\ref{gbs}).

The scattered photon flux (Fig.~\ref{fig:flux}) was calculated using the 
following formulae:
\begin{equation}
{{dN_\gamma} \over {ds}} = {{dL_{e\gamma}}\over{ds}} \cdot \sigma_c.
\end{equation} 
The total yield of photons is about 17000 ph/s/mA/W 

\subsection{Location of optic elements}

\begin{figure}[p]
\centering
\includegraphics*[width=10cm,height=6.5cm]{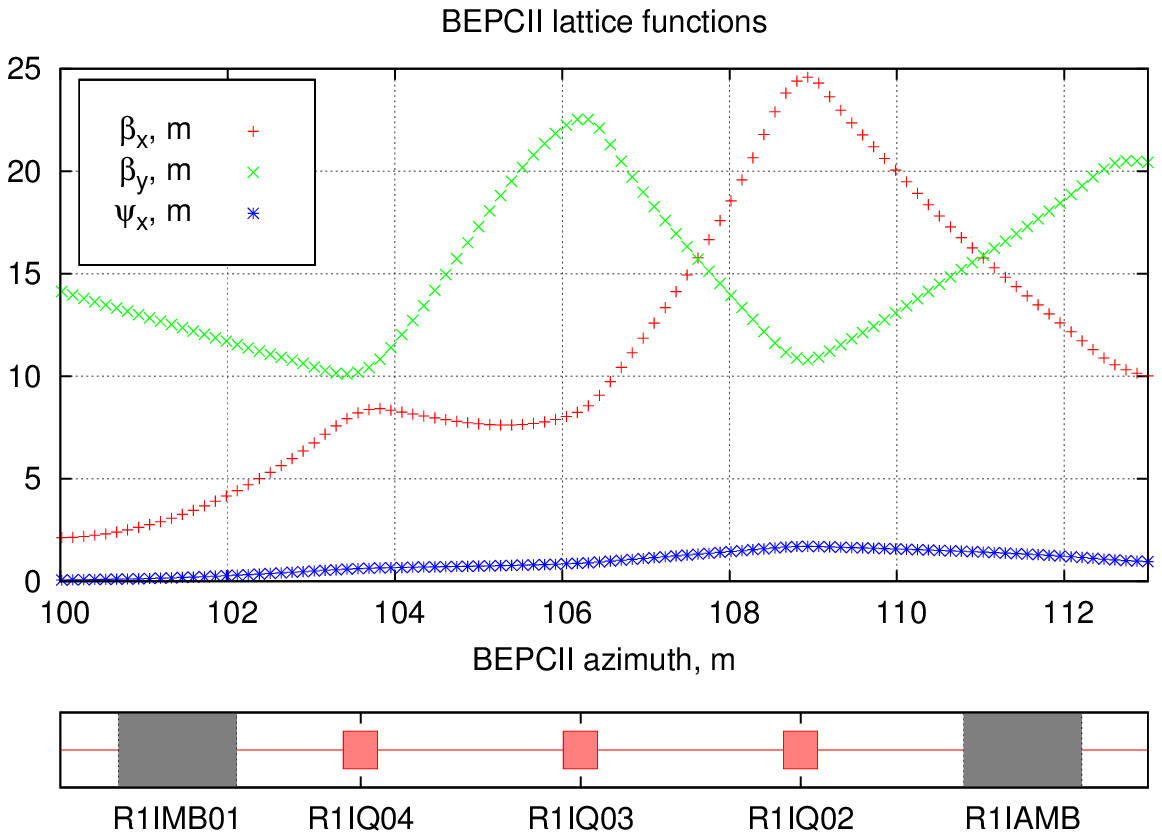}
\caption{\em BEPC-II lattice functions in the interaction area. The positions 
of collider's magnets are shown at the bottom of the picture.}
\label{fig:lf}
\includegraphics*[width=10cm,height=6.5cm]{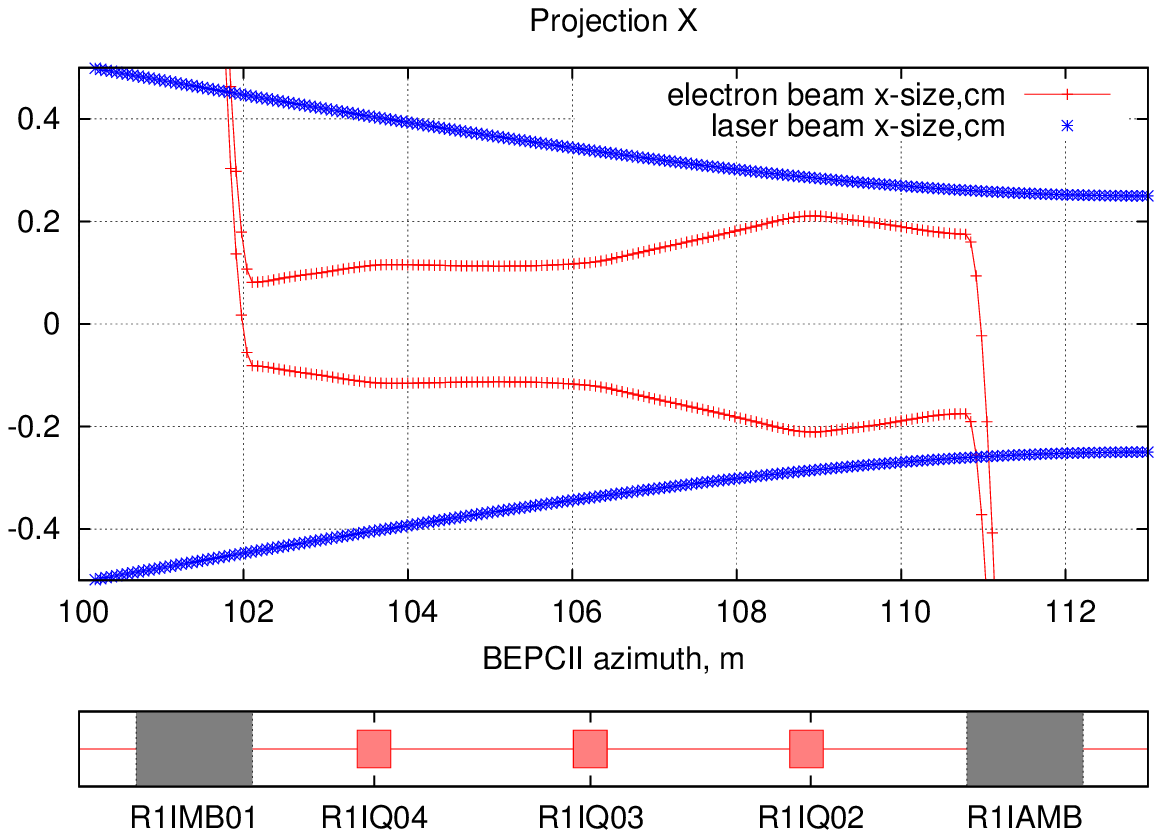}
\caption{Laser and $e^-$ or $e^+$ beams size ($\pm\sigma$) in horizontal 
plane.}
\label{fig:x}
\end{figure}

\begin{figure}[p]
\centering
\includegraphics*[width=10cm,height=6.5cm]{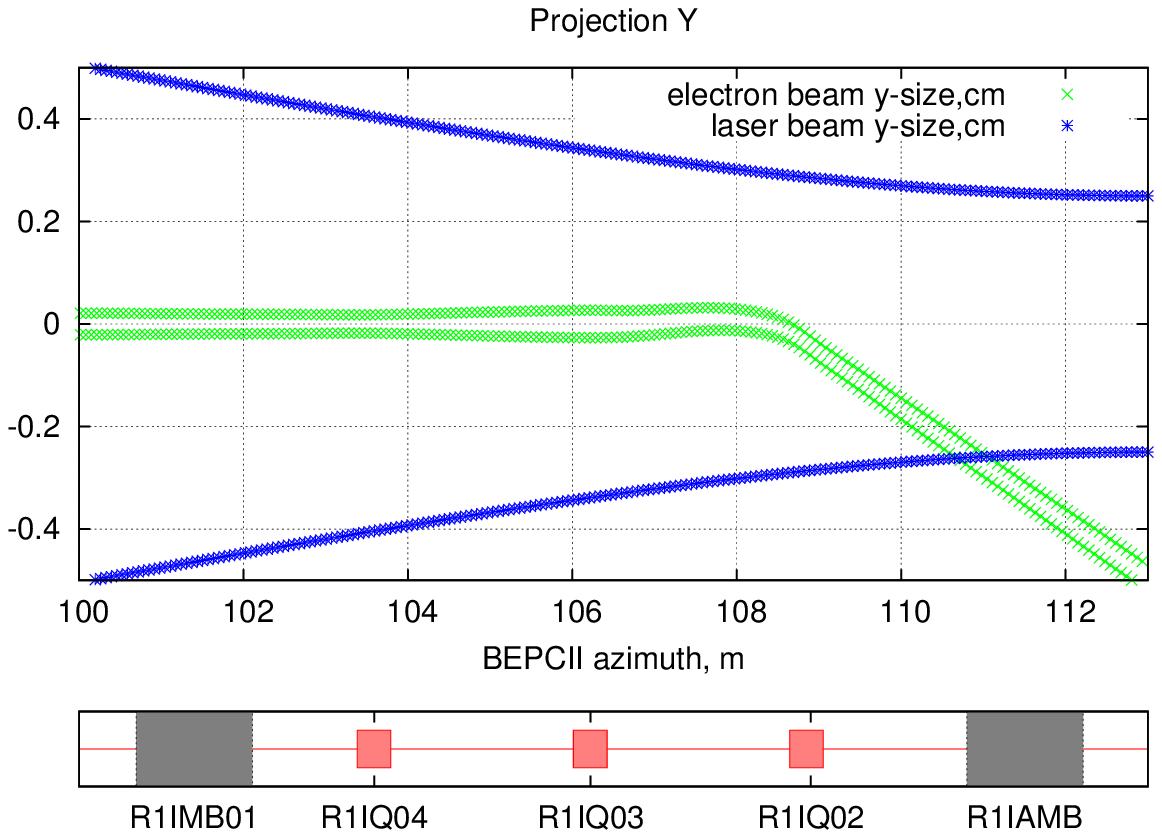}
\caption{The Laser and $e^-$ or $e^+$ beams size ($\pm\sigma$) in vertical 
plane. The vertical bending for the $e^-$ and $e^+$ has the opposite sign.}
\label{fig:y}
\includegraphics*[width=10cm,height=6.5cm]{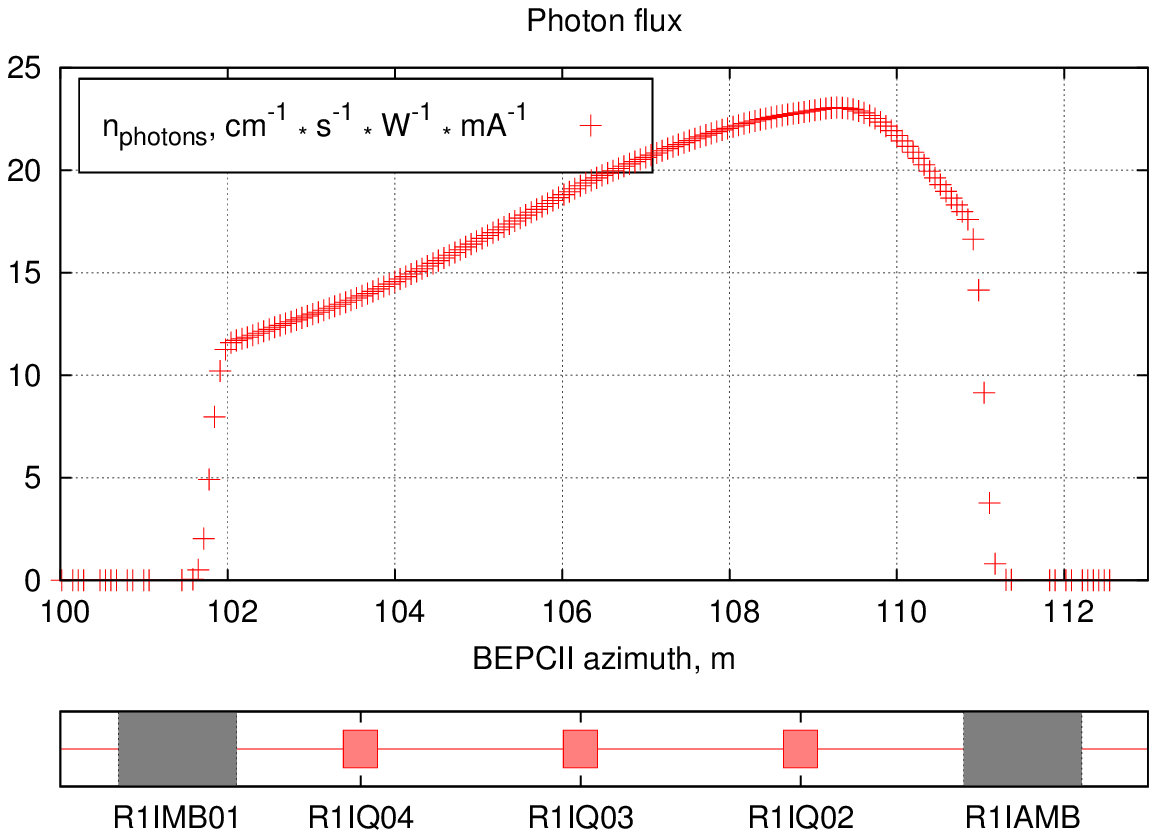}
\caption{The scattered photons flux. The beam energy is $\varepsilon=1890$~MeV,
beam current $I_e=1$~mA, laser power $P=1$~W. The positions of collider
magnets are shown at the bottom of the picture.}
\label{fig:flux}
\end{figure}

Optic components require careful alignment and mechanical stability. It is 
very convenient to have a free access to them independently from the collider 
operation. This elements should be placed outside the BEPC-II tunnel in the 
neighbour hall (which is available and empty now) as shown in 
Fig.~\ref{fig:layout}. 
This requires drilling a holes in the inner wall of the BEPC-II tunnel. The 
following optical units are situated along the collider tunnel wall:
\begin{enumerate}
\item 
the mobile reflection prism which provide to change the direction of laser 
beam insertion (to e$^+$ or e$^-$ beam pipe),
\item
the two mirrors which provides the laser beam turning on the angle 90$^\circ$. 
The mirrors are installed on a special supports with precise vertical and 
horizontal angular alignment by stepping motors (one step equals to 
$1.5\times 10^{-6}$~rad).
\end{enumerate}
The other optical elements: two lenses with the focal length $f=40$~cm and 
laser are placed along the neighbour wall 
(Fig.~\ref{fig:layout}).  The total distance from the laser output aperture to
the entrance flange of the BEPC-II vacuum chamber is about 18.0~m. The laser 
beam transverse size at this flange should be equal to 0.20 -- 0.25~cm 
(Fig.\ref{fig:x},~\ref{fig:y}). The lenses should be placed at distance 300.0 
and 381.6 cm from the laser in order to provide this size (Fig.\ref{fig:flux}).
\begin{figure}[t]
\centering
\includegraphics*[width=0.75\textwidth]{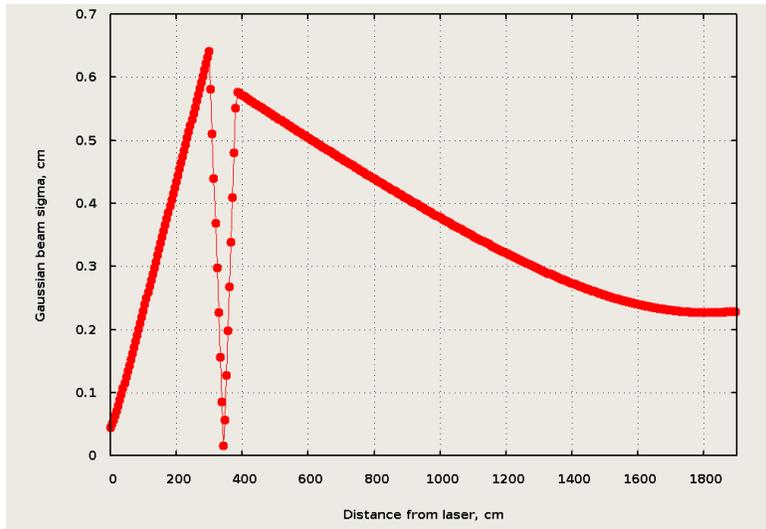}
\caption{The laser beam transverse size versus the distance from laser 
position (at the origin) along the optical axis. The lenses are situated at 
distance 300.0 and 381.6 cm from the laser position, the entrance flange of 
the BEPC-II vacuum chambers is at the distance of 18~m from the laser 
position.}
\label{fig:lensp}
\end{figure}

\section{Conclusion}
The proposed energy calibration system will provide relative accuracy 
$\delta\varepsilon / \varepsilon \leq 3\times 10^{-5}$ in the beam energy 
range from 1 to 2 GeV. Some of system units are ready or in work. The goal of 
the further efforts should be the manufacture and purchase of the system 
elements and  its installation at launching at BEPC-II to January 2009.
\section{Acknowledgments}
The authors are grateful to Fred Harris, Wolfgang Kuehn, Qian Liu, 
Stephen Olsen, QingBiao Wu, JianYong Zhang and QingJiang Zhang for useful 
discussions.

\end{document}